\documentclass[showpacs,preprint,amsmath,superscriptaddress,prl]{revtex4}  % alvaro modified
\usepackage[dvips]{graphicx}
\usepackage{epsfig}
\usepackage{color}
\usepackage{subfigure}
\usepackage{soul}

\begin{document} 
\title{
Phase transition,
scaling of moments, and order-parameter
distributions in 
%%Galton-Watson 
%first-%
%return-time distributions of 
Brownian particles
and
branching processes
with finite-size effects
} 
\author{\'Alvaro Corral}
\affiliation{
Centre de Recerca Matem\`atica,
Edifici C, Campus Bellaterra,
E-08193 Barcelona, Spain
}\affiliation{Departament de Matem\`atiques,
Facultat de Ci\`encies,
Universitat Aut\`onoma de Barcelona,
E-08193 Barcelona, Spain
}\affiliation{Barcelona Graduate School of Mathematics, 
Edifici C, Campus Bellaterra,
E-08193 Barcelona, Spain
}\affiliation{Complexity Science Hub Vienna,
Josefst\"adter Stra$\beta$e 39,
1080 Vienna,
Austria
}
\author{Rosalba Garcia-Millan}
\affiliation{
Department of Mathematics, Imperial College London, 
180 Queen's Gate, London SW7 2AZ, UK}
\author{Nicholas R. Moloney}
\affiliation{London Mathematical Laboratory, 
14 Buckingham Street, London WC2N 6DF, UK}
\author{Francesc Font-Clos}
\affiliation{
Center for Complexity and Biosystems, Department of Physics, University of Milan, via Celoria 16, 20133 Milano, Italy}
%\author{\'Alvaro Corral}
%% 
%%\email{acorral@crm.es} 
%\affiliation{Centre de Recerca Matem\`atica,
%Edifici C, Campus Bellaterra,
%E-08193 Barcelona, Spain.
%} 
%\affiliation{Departament de Matem\`atiques,
%Facultat de Ci\`encies,
%Universitat Aut\`onoma de Barcelona,
%E-08193 Barcelona, Spain}
% 
% 
\begin{abstract} 
We revisit the problem of Brownian diffusion with drift in order to
study finite-size effects in the geometric Galton-Watson branching
process. This is possible because of an exact mapping between
one-dimensional random walks and geometric branching processes, known
as the Harris walk.  In this way, first-passage times of Brownian
particles are equivalent to sizes of trees in the branching process
(up to a factor of proportionality).  Brownian particles that reach a
distant boundary correspond to percolating trees, and those that do
not correspond to non-percolating trees.  In fact, both systems
display a second-order phase transition between ``insulating'' and
``conducting'' phases, controlled by the drift velocity in the
Brownian system.  In the limit of large system size, we obtain exact
expressions for the Laplace transforms of the probability
distributions and their first and second moments.  These quantities
are also shown to obey finite-size scaling laws.
\end{abstract} 
% \pacs{64.60.av, 91.30.-f, 81.30.Kf,05.50.+q}

\date{\today}

\maketitle

% empezado 18/8/2016

%OJO!!! El expo 3/2 no vale tambien para Pe pequenyo??
%Y decir que vale tambien para $f_r$ !!!

%{\small
%Escape: se escapa: many times!!

%First return:  3 times, including title

%first return: va y vuelve. OJO, para $x=0$ tambien puede ser escape
%(si $x_0$ fuera exactamente 0)

%Duration of the walk: 2 times
%
%Passage: 0 times
%
%Path??? Excursions??? 0 times both
%
%Trees?? Many times!! Clusters?? Several times Realizations??? 3 times 
%}

%TREES versus CLUsTER??????

%VERTICES wins to SITES and nodes

%$\ell$ finite and infinite???

%\newpage

\section{Introduction}
\begin{figure}[ht]
%\hspace{-3cm}
\includegraphics[width=1.\columnwidth]{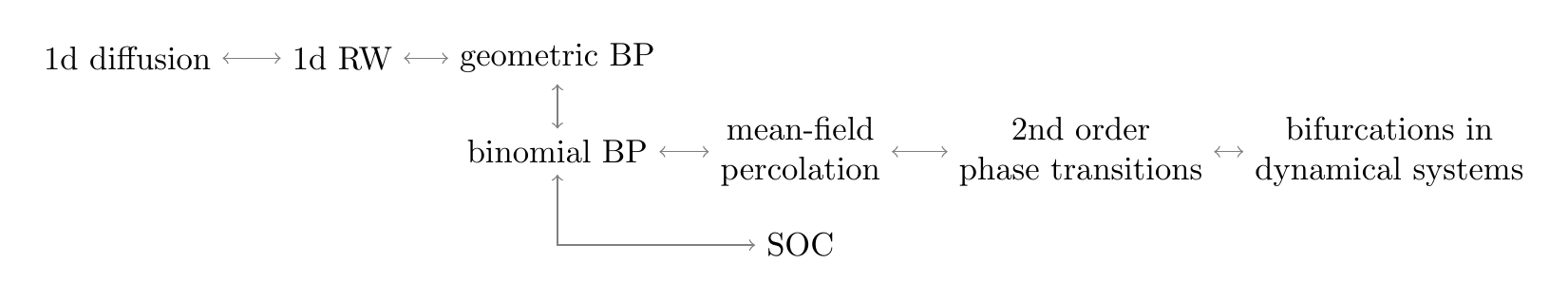}\\
%\includegraphics[width=.80\columnwidth]{fig2.pdf}\\
%\vspace{-10cm}
\caption{
Relations between different 
%statistical-physics phenomena.
physical processes and models, with abbreviations
RW (random walk), BP (Galton-Watson branching process),
%\cite{Harris_original, branching_biology}, 
SOC (self-organized critical phenomena)
%\cite{Jensen,Christensen_Moloney,Pruessner_book}.
\cite{Jensen,Christensen_Moloney, Watkins_25years}.
BP with geometric and binomial offspring distribution share the same
critical properties due to universality.  }
\label{Fig1}
\end{figure}

Random walks and rooted trees are important models in
probability theory and statistical physics \cite{Harris52}.  Random
walks \cite{Weiss_Rubin,Weiss_RW,Klafter_Sokolov} provide a
microscopic model for diffusion processes \cite{Crank_diffusion}, and
rooted trees are the geometric representation of branching processes
\cite{Harris_original,branching_biology}.  The Galton-Watson process,
which is the simplest branching process and is at the heart of
self-organized-critical behavior
\cite{Zapperi_branching,Corral_FontClos}, is essentially the same as
mean-field percolation \cite{Aharony,Christensen_Moloney} when the
offspring distribution of the former is binomial.  Percolation,
meanwhile, is one of the simplest examples of a second-order phase transition
\cite{Stanley,Yeomans1992,munoz_colloquium}.  Moreover, important
characteristics of phase transitions also show up in bifurcations in
low-dimensional dynamical systems \cite{Corral_Alseda_Sardanyes}.  An
exact, purely geometric mapping between walks and rooted trees was
presented in Ref. \cite{Harris52}, thereby connecting
results from random walks to branching processes and beyond
\cite{Font_clos_molon}.  Figure \ref{Fig1} offers a scheme for these
relations.

%EQUIVALENTE A SITE O A BOND PERCO??
% es lo mismo en la red de Bethe

This mapping, known as the Harris walk, implies that a realization of
a finite-size geometric Galton-Watson branching process with no more
than $L$ generations corresponds (exactly) to a random walker confined
between absorbing and reflecting boundaries (at $X=0$ and $X=L$,
respectively). Recently, Font-Clos and Moloney \cite{Font_clos_molon}
applied this mapping to derive the distribution of the size of the
percolating clusters in a finite Bethe lattice, by using the
first-passage time to the origin of a Brownian particle conditioned to
first reach $L$.  In the critical case (unbiased diffusion), a
Kolmogorov-Smirnov distribution is obtained (as in Ref.~\cite{Botet}).
In the subcritical case (diffusion with negative drift) these authors
find that the size of the percolating cluster follows Gaussian
statistics, whereas in the supercritical case (diffusion with positive
drift), an exponential-like distribution is reported. 
% In this work we follow the same approach as in Ref. \cite{Font_clos_molon},
% using the first return-time in a diffusion process to calculate not
% only the distribution of the size of a percolating cluster but the
% size distribution of any cluster (percolating or not) in a geometric
% Galton-Watson process, when finite-size effects are considered (note
% that, in this context, clusters in percolation are totally equivalent
% to Galton-Watson trees).

In this work we follow the approach of Ref.~\cite{Font_clos_molon} and use 
the first passage-time in a diffusion process to calculate
the size distribution of both percolating and non-percolating clusters
in a geometric Galton-Watson process with a finite number of generations.
%
%%Y que mas???
%
%Whereas a percolating cluster corresponds to a random walker that reaches the
%$X=0$ boundary after touching the reflecting boundary (at $X=L $),
%a non-percolating cluster corresponds to a walker that reaches $X=0$
%without touching the other boundary.
First, we solve the corresponding diffusion problem and derive
analytical expressions and scaling laws (Sec. 2); then we translate
the results to the random-walk picture (Sec. 3); and finally, by means
of the mapping from trees to walks \cite{Harris52}, we obtain the
properties of the associated branching process.  A discussion about
the most appropriate definition of an order parameter in these systems
is also provided.  Our main focus is the size distribution of all
clusters (whether they percolate or not). Given that the percolating
case was thoroughly studied in \cite{Font_clos_molon}, we will only
provide details for the non-percolating case.
% This is in the same way as Font-Clos and Moloney \cite{Font_clos_molon},
% who descomposed 
% the calculation of the first-passage time after touching the reflecting boundary
% into the independent sum of the time to reach a virtual absorbing boundary
% at $X=L$ plus the time to go from the real reflecting boundary at $X=L$ to the absorbing 
% one at $X=0$.
%And as the other boundary is not reached, we can replace its reflecting character
%by an absorbing one, for simplicity.

\section{First-passage times via the diffusion equation}

Consider the one-dimensional diffusion equation with drift,
\begin{equation}
\frac{\partial c}{\partial t} + v \frac{\partial c}{\partial x} =
D \frac{\partial^2 c}{\partial x^2}
\label{diffeq}
\end{equation}
which describes the evolution of the concentration $c(x,t)$ of
particles at position $x$ and time $t$, with drift velocity $v$ and
diffusion constant $D$.  Both position and time are continuous.  We
will work in a finite interval, $0\le x\le \ell$, with $\ell$ playing
the role of system size ($X$ and $L$ mentioned in the introduction are
dimensionless versions of $x$ and $\ell$).  Following
Redner~\cite{Redner}, first-passage times are most readily obtained by
applying the Laplace transform $c(x,s)=\int_0^\infty e^{-s t} c(x,t)
dt$ to the diffusion equation, yielding
\begin{equation}
%s c(x,s)-\delta(x-x_0) +v c'(x,s) = D c''(x,s),
s c(x,s)-c(x,t=0) +v c'(x,s) = D c''(x,s),
\label{diffeqlaplace}
\end{equation}
where the prime denotes a derivative with respect to $x$.

\subsection{Absorbing boundaries}

First-passage time probability densities $f(t)$ are obtained from
spatial concentration gradients at absorbing boundaries~\cite{Redner}.
To see this, we track the rate of particle loss from the interval
$0\le x\le \ell$:
\begin{align}
f(t) = -\frac{d}{dt} \int_0^{\ell} c(x,t) dx  &= 
-
\int_0^{\ell} D
\frac{\partial^2 c}{\partial x^2} \, dx 
+ %- 
\int_0^{\ell} v \frac{\partial
  c}{\partial x} \, dx \\
 &= \left. -D \frac{\partial c}{\partial x} \right|_{x= \ell}
+ \left. D \frac{\partial c}{\partial x} \right|_{x=0},
\end{align}
where we have made use of 
the normalization of $c(x,t)$ for $t=0$
and of
the absorbing boundary conditions at $x=0$
and $x=\ell$:
\begin{equation}
c(x=0,t) = c(x=\ell,t) = 0.
\label{abc}
\end{equation}
Note that the terms in the above sum above are not probability
densities themselves (because they are not normalized).  Rather, they
are the net outflux of particles at each boundary, so that
$f(t)=j_\ell(t)+j_0(t)$.  The Laplace transform of the probability
density can therefore be written as
\begin{equation}
f(s)=j_\ell(s)+j_0(s)=-D c'(x= \ell,s) + D c'(x=0,s).
\label{jLj0}
\end{equation}

With Dirac-$\delta$ initial condition centered at $x_0$, 
\begin{equation}
c(x,t=0)=\delta(x-x_0),
\label{deltainitial}
\end{equation}
the solution of the Laplace-transformed diffusion equation with two
absorbing boundaries is~\cite{Redner}: 
\begin{equation}
c(x,s)=
%\exp\left[ P_e\left( \frac x L - {u_0} \right) \right]
%e^{P_e(x/ \ell- {u_0})}
\frac{e^{(x/ \ell- {u_0})P_e} \sinh(Rx_</\ell) \sinh[R( \ell-x_>)/\ell]}{D R\sinh(R)/\ell},
\label{lagorda}
\end{equation}
where $P_e$ is a dimensionless parameter known as the P\'eclet number
(up to a factor of $1/2$ according to convention),
\[
P_e=\frac{\ell v}{2D},
\]
$u_0$ is the dimensionless initial position and $\tau$ a diffusion time,
\[
  {u_0}=\frac {x_0} \ell, \qquad \tau=\frac{\ell ^2}D,
\]
and, for convenience of notation,
\[
R=\sqrt{P_e^2+ \tau s},\qquad\text{and}\qquad
x_< =\min(x,x_0), \, x_>=\max(x,x_0).
\]

\subsection{Absorption at $x=0$}

\begin{figure}[ht]
\includegraphics[width=.99\columnwidth]{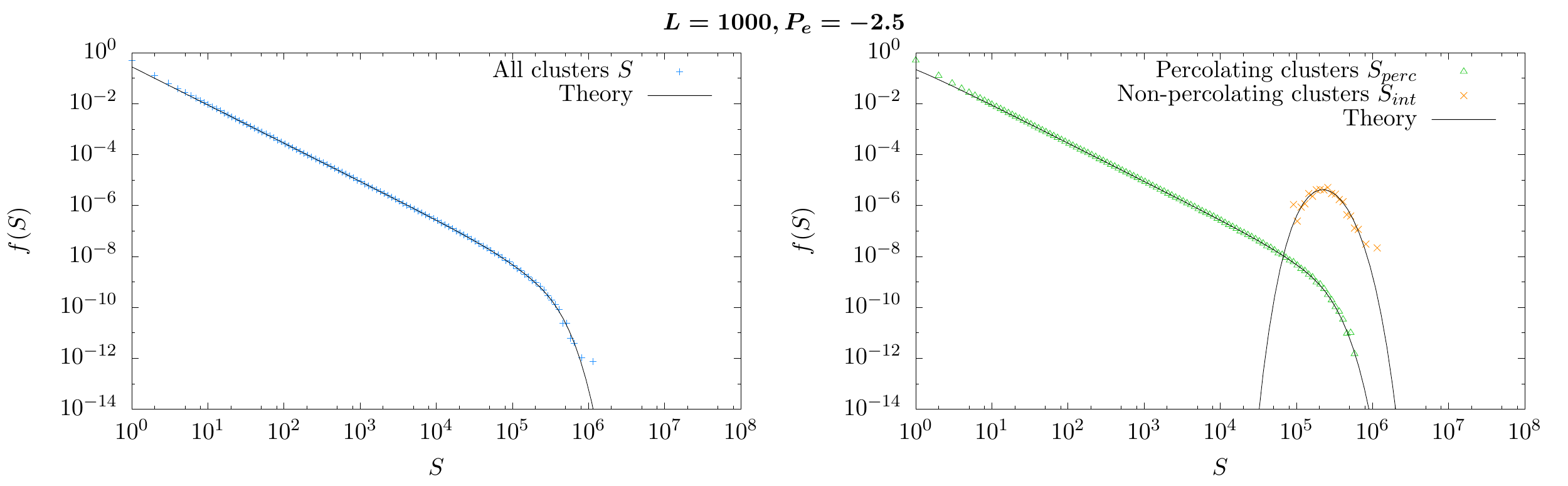}\\
\includegraphics[width=.99\columnwidth]{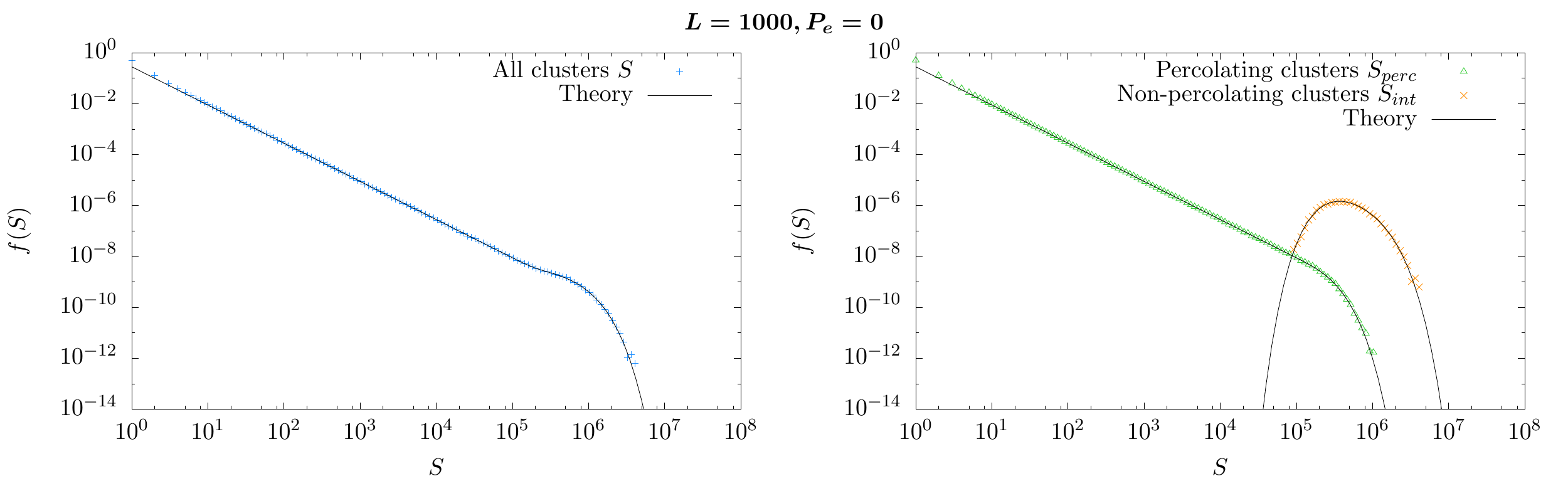}\\
\includegraphics[width=.99\columnwidth]{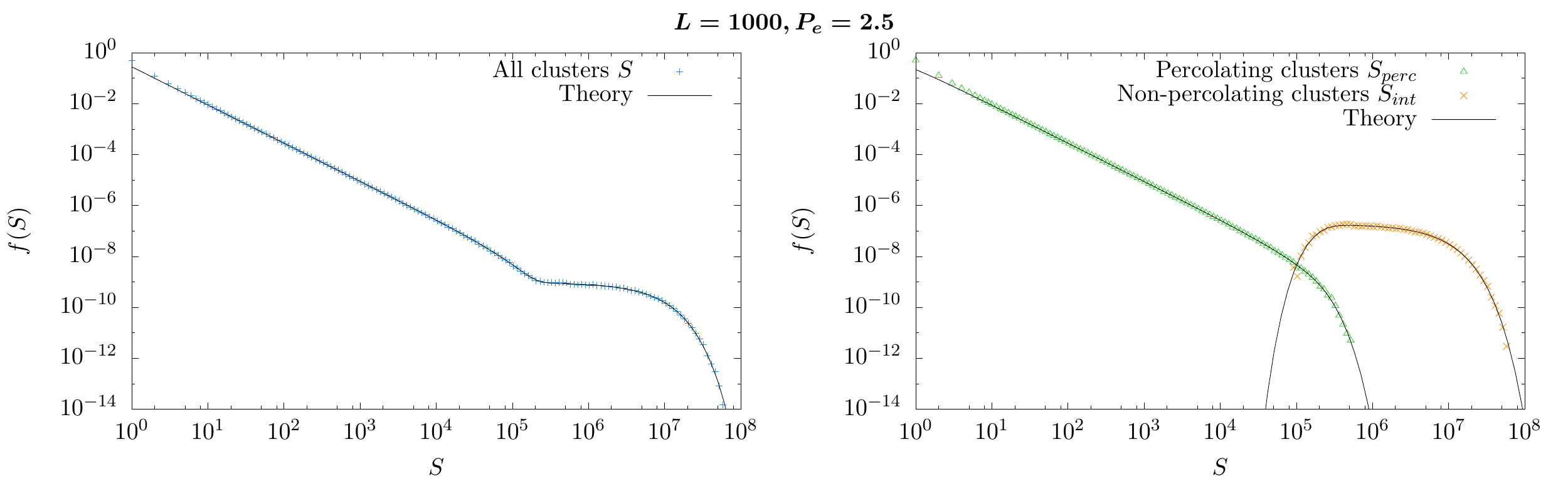}\\
\caption{ Distribution of sizes in the geometric Galton-Watson process
  in the three regimes: subcritical ($P_e=-2.5$, top), critical
  ($P_e=0$, middle), and supercritical ($P_e=2.5$, bottom), together
  with theoretical predictions from the first-passage times of a
  diffusion process
(numerical inversion of the Laplace transform is done by the Talbot method using the 
routine in Ref. \cite{Laplace_inversion}).  Plots on the left show the full
  distribution. Plots on the right show the distribution decomposed
  into its percolating and non-percolating contributions.  In the
  critical regime, the complete distribution $f(S)$ (middle left) shows
  a small bump for large sizes, which is much more pronounced in the
  supercritical regime.  The subcritical distribution, having no bump
  (top left), can be visually confused with a critical distribution.
  Logarithmic binning has been
  applied~\cite{Christensen_Moloney,Corral_Deluca}.  }
\label{Fig_distributions}
\end{figure}

Using the above formalism, Redner \cite{Redner} examines the driftless
case, $v=0$, in full detail. For completeness, we provide the
calculation for $v\neq 0$ in Appendix A. In summary, the
Laplace transform of the first-passage time density, $f_0(s)$, for
absorption at $x=0$ can be expanded in powers of $u_0$ as
\begin{equation}
f_0(s)=
%%\frac{j_0(s)}{j_0(s=0)}=
1-
\left(\frac R {\tanh R} - \frac {P_e} {\tanh P_e} \right) {u_0} 
+ \mathcal{O}({u_0}^2).
\label{nuevanueva_noapp}
\end{equation}
A numerical inversion of $f_0(s)$
is shown in Fig. \ref{Fig_distributions}.

\begin{figure}[ht]
\includegraphics[width=.99\columnwidth]{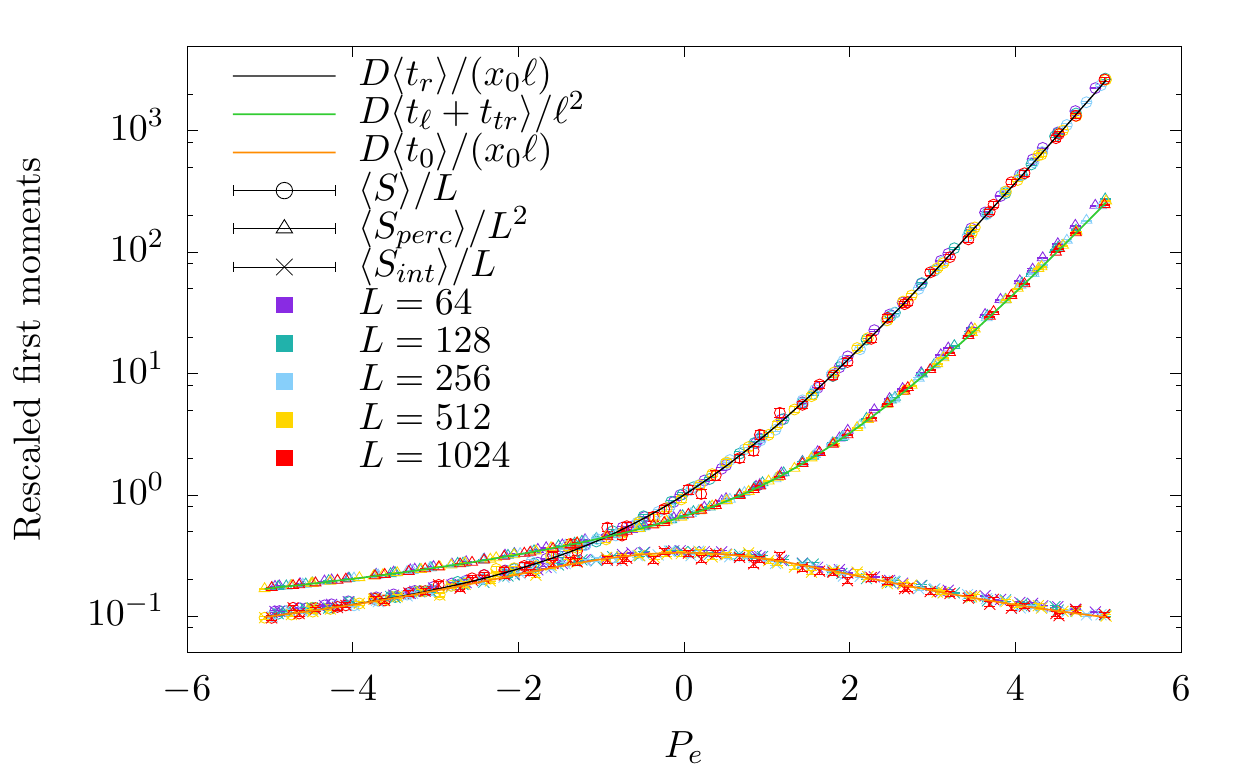}
\caption{Rescaled mean sizes (simulations, $10^5$ realizations) and rescaled times
  (theory) versus P\'eclet number for different systems sizes $L$.}
\label{Fig_moments}
\end{figure}

The first two moments can be similarly expanded as
\begin{align}
\langle t_0 \rangle &=
\frac{1}{2 P_e \tanh P_e} \left(1 + P_e \tanh P_e  - \frac {P_e}{\tanh P_e}\right)\tau {u_0}
+ \mathcal{O}({u_0}^2),
\label{t0_noapp} \\
\langle t_0^2 \rangle &=
\frac{1}{2 P_e^3\tanh P_e}\left( 
\frac 1 2
 - \frac {P_e \tanh P_e} 2  + \frac{P_e}{2 \tanh P_e} +P_e^2- \frac{P_e^2}{\tanh^2 P_e} 
\right) {\tau^2{u_0}}  + \mathcal{O}({u_0}^2).
\label{t02_noapp}
\end{align}
%Note that the
%factor $\tau u_0$ can also be written as $\tau
%u_0=x_0\sqrt{\tau/D}=x_0 \ell /D$, whereas $\tau^2
%u_0=x_0\sqrt{\tau^3/D}=x_0 \ell^3/ D^2$.
A plot of $\langle t_0 \rangle$ as a function of $P_e$ is shown in Fig. \ref{Fig_moments}.

Note that these expressions are even in $P_e$. The expansions are
valid in the limit of small $u_0$, i.e., $x_0 \ll \ell$ (more
precisely, $u_0 P_e \ll 1$ and $u_0 R \ll 1$).  See Eqs.~(\ref{t0}),
(\ref{t02}), and (\ref{nuevanueva}) in Appendix A for their full
derivation.
%An exact alternative to Eq. (\ref{nuevanueva_noapp}) is
%$$%\begin{equation}
%f_0(s)=
%%%%\frac{j_0(s)}{j_0(s=0)}=
%\frac
%{\sinh[(1-u_0)R]    \sinh P_e}
%{\sinh[(1-u_0)P_e] \sinh R}.
%%\label{nuevayexacta}
%$$%\end{equation}

\subsection{Critical point}

The critical point corresponds to diffusion with no drift, $P_e=0$
(i.e. $v=0$). Using the results of Appendix A, the exact distribution
reads
\begin{equation*}
f_0^*(s) =\frac{\sinh[(1-u_0)\sqrt{\tau s}]}{(1-u_0)\sinh\sqrt{\tau s}},
\end{equation*}
where the asterisk denotes the critical point. To first order in $u_0$
\begin{equation}
  f_0^*(s)=
1-\left(\frac { \sqrt{\tau s}} {\tanh \sqrt{\tau s}} -1\right) u_0
+\mathcal{O}({u_0}^2).
\label{otramasymas}
\end{equation}
Figure \ref{Fig_distributions} shows a plot of the numerical inverse
Laplace transform of $f_0^*(s)$.  The case $P_e=0$ and $\ell$ infinite
corresponds to $\tau s \gg 1$, for which
\begin{equation*}
  f_0^*(s)=
  j_0^*(s)\simeq 1- { {u_0}\sqrt{\tau s}} +  \mathcal{O}({u_0}^2).
\end{equation*}
After inverting the Laplace transform (see e.g. Eq. 4.6.23 of
Ref. \cite{Bleistein}), 
\begin{equation}
f_0^*(t)=
j_0^*(t) \simeq\frac {{u_0}} 2 \sqrt{\frac \tau {\pi t^3}} 
+ \mathcal{O}({u_0}^2),
\label{powerlaw32}
\end{equation}
for $P_e=0$ and $t \ll \tau=\ell ^2/D$, thereby recovering the
well-known $t^{-3/2}$ behavior of first-passage to the origin in an
infinite system.

To lowest order in $u_0$, the moments are given by
\begin{align}
\langle t_0^* \rangle &=\frac {\tau {u_0}} 3 = \frac{x_0
  \ell}{3D}, \label{t0ave} \\
\sigma_0^*\simeq \sqrt{\langle t_0^{2*}  \rangle} &=\sqrt{\frac {2 \tau^2 {u_0}}{45}} = \sqrt{\frac{2 x_0 \ell ^3}{45 D^2}},
\label{sigma0}
\end{align}
expanding Eq. (\ref{otramasymas})
for small $\sqrt{\tau s}= \ell\sqrt{s/D}$. 
%Let us note that the expressions for
%both moments, Eqs. (\ref{t0_noapp})-(\ref{t02_noapp}), are even
%in $P_e$ (in the same way as Eq. (\ref{nuevanueva_noapp})).

%In principle, the expansion we have performed is not valid for $P_e=0$
%(because $z=\tau s /P_e^2$ has to be small), but we obtain the same
%result turning back to Eq. (\ref{firstordeninell}) in Appendix A
%(before the expansion in $z$) and taking $P_e=0$ there.  This leads to
%$$%\begin{equation} 
%j_0^*(s)=1-\frac { {u_0}\sqrt{\tau s}} {\tanh \sqrt{\tau s}}
%+\mathcal{O}({u_0}^2),
%$$%\end{equation}
%NEW EQUATION!!!
%and

%from where the same expressions (\ref{t0ave}) and (\ref{sigma0})
%for ${\langle t_0^* \rangle}$ and $\sigma_0^*$
%={\langle t_0^2 \rangle}$ 
%are obtained
%in the case of 
%$\ell$ finite, corresponding to 
%$\sqrt{\tau s}= \ell\sqrt{s/D}$ small.  In any case, the result
%(\ref{t0ave}) for $\langle t_0^* \rangle$ is the same as the one
%obtained in Ref. \cite{Redner}.

%
%The case of $P_e=0$ and $\ell$ infinite corresponds to $\tau s \gg 1$, 
%which leads to 
%$$%\begin{equation} 
%f_0^*(s)=
%j_0^*(s)\simeq 1- { {u_0}\sqrt{\tau s}} +  \mathcal{O}({u_0}^2),
%$$%\end{equation}
%containing also terms of order higher than $\sqrt{s}$.  The equation
%for $f_0^*(s)$ is the Laplace transform of a function with a power-law
%tail, 

%CARE, IN THE 2 EQS ABOVE, 
%HIGHER TERMS $\mathcal{O}(s)$ ARE MISSING!!!
%%OJO aqui, distinguir L finito y L infinito!!!!

\subsection{Absorbing boundary at $x=\ell$ and splitting probability}

Although already studied in Ref. \cite{Font_clos_molon}, for
completeness we summarize the results for absorption at $x=\ell$. In
the notation of this article, we extract from Eq. (\ref{lagorda}) the
expansion
\begin{equation}
j_\ell(s)=-D c'(x= \ell,s)= \frac{e^{(1-{u_0})P_e} \sinh({u_0} R)}{\sinh R}
%=
%\begin{equation}
%\end{equation}
= \frac{e^{P_e} R {u_0}}{\sinh R} + \mathcal{O}({u_0}^2),
\label{jL}
\end{equation}
valid for $u_0 \ll 1$. For $z=\tau s /P_e^2 \ll 1$,
\begin{equation}
j_\ell(s)
=\frac{e^{P_e} P_e {u_0}}{\sinh P_e} 
\left[1-\left(-1 +\frac{P_e}{\tanh P_e} \right) \frac z 2  + \frac 1 2
 \left( \frac{2 P_e^2}{\tanh^2 P_e} - \frac{P_e}{\tanh P_e} -P_e^2 -1 \right) \frac {z^2} 4 + \mathcal{O}(z^3)\right]
+ \mathcal{O}({u_0}^2),
\label{jLlin}
\end{equation}
So, to zeroth order in ${u_0}$,
\begin{align}
\langle t_\ell \rangle &=\frac \tau {2 P_e^2} \left(\frac{P_e}{\tanh P_e} -1\right)
+ \mathcal{O}({u_0}),
\label{tL} \\
\langle t_\ell ^2 \rangle &= \frac {\tau^2} {4 P_e^4} 
\left(\frac{2 P_e^2}{\tanh^2 P_e} - \frac{P_e}{\tanh P_e} -P_e^2 -1\right)
+ \mathcal{O}({u_0}).
\label{tL2}
\end{align}
%On the other hand,
%taking $P_e=0$ 
%%and $\ell$ finite ??? 
%in Eq. (\ref{jL}),
%\begin{equation}
%j_\ell^*(s)= 
%\frac{\sinh[u_0\sqrt{\tau s}]}{\sinh{\sqrt{\tau s}}}
%=\frac{ {u_0} \sqrt{\tau s}}{\sinh \sqrt{\tau s}}
%+ \mathcal{O}({u_0}^2),
%\end{equation}
%which is the Laplace transform of the celebrated Kolmogorov-Smirnov distribution \cite{Botet,Font_clos_molon}
%(except for the unnormalizing factor ${u_0}$).
%
%%And for $P_e=0$ and $\ell$ infinite?????
%%COMPROBAR QUE LOS MOMENTOS PARA $P_e=0$ DAN LO MISMO
%%DE AMBAS MANERAS!!
%%
%%Ojo, decir que $\frac{e^{P_e} P_e {u_0}}{\sinh P_e} $ es lo mismo que en Refs. \cite{GarciaMillan,Corral_garciamillan}
%%(y en plos one!!)
%%con $P_e\simeq 2\ell(q-1/2) = y/2$
%%porque $v=2q-1$ y $D=2pq$
%%(cogiendo unidad de esp y tiempo 1)
%%y tambien $x_0=1$

Note that the common multiplying factor in Eq. (\ref{jLlin}),
\begin{equation}
C_\ell=
j_\ell(0)=\frac{e^{(1-{u_0})P_e} \sinh({u_0} P_e)}{\sinh P_e}=
\frac{e^{P_e} P_e {u_0}}{\sinh P_e}
+ \mathcal{O}({u_0}^2).
\label{CL}
\end{equation}
gives the ratio between the outflux of particles at $\ell$, denoted by
$j_\ell(t)$, and the probability density $f_\ell(t)$ of the
first-passage time to the boundary at $x= \ell$; i.e.,
$j_\ell(t) = C_\ell f_\ell(t)$, so that,
$$
f_\ell(s)=\frac{j_\ell(s)}{C_\ell}=
\frac {\sinh P_e}{\sinh R} 
\frac {\sinh (u_0 R)}{\sinh (u_0 P_e)}=
\frac{R \sinh P_e}{P_e \sinh R}+\mathcal{O}(u_0).
$$ 
$C_\ell$ is known as the splitting probability \cite{Redner,Farkas}.
Note that $\langle t_\ell \rangle$, $\langle t_\ell^2 \rangle$,
and $f_\ell(s)$ (and therefore $f_\ell(t)$) are even for $P_e$
(but not $j_\ell(s)$ and $C_\ell$).
%As $C_\ell=C_\ell^*=u_0$ 
For $P_e=0$ (i.e., at the critical point) one has that $f_\ell^*(s)$, to
first order in $u_0$, is the Laplace transform of the celebrated
Kolmogorov-Smirnov distribution \cite{Botet,Font_clos_molon},
\begin{equation}
f_\ell^*(s)= 
\frac {\sinh (u_0 \sqrt{\tau s})}{u_0 \sinh \sqrt{\tau s}}=
\frac{\sqrt{\tau s}}{\sinh \sqrt{\tau s}}
+ \mathcal{O}({u_0}).
\label{bis21}
\end{equation}

The factor $C_\ell$ is also the probability that a particle is
absorbed at $x= \ell$.  Equation (\ref{CL}) is
the same scaling law found in
Refs. \cite{GarciaMillan,Corral_garciamillan}, as explained in
the next section.  A formula for $C_\ell$ also appears in
Refs. \cite{Redner,Farkas,Font_clos_molon}, but without expanding in ${u_0}$,
therefore hiding its finite-size scaling. The probability that a
particle is instead absorbed at $x=0$ is $C_0=1-C_\ell \simeq 1$ (to zeroth order in
$u_0$).
%Finally, the exact counterparts for $C_\ell$ and $f_\ell(s)$ above are
%$$
%C_\ell=
%j_\ell(0)=\frac{e^{(1-{u_0})P_e} \sinh({u_0} P_e)}{\sinh P_e}
%$$
%and
%$$f_\ell(s)=
%\frac {\sinh P_e}{\sinh R} 
%\frac {\sinh (u_0 R)}{\sinh (u_0 P_e)},
%$$
%which at the critical point ($P_e=0$) becomes
%$$
%f^*_\ell(s)=
%\frac {\sinh (u_0 \sqrt{\tau s})}{u_0 \sinh \sqrt{\tau s}}.
%$$

Combining the results for $t_\ell$ with those obtained previously for
$t_0$, we have the solution of one-dimensional diffusion between two
absorbing boundaries. In Appendix B, we show that it displays a phase
transition with finite-size scaling when $x_0 \ll \ell$~\cite{Farkas}.

%A formula for $C_\ell$
%also appears in Refs. \cite{Redner,Font_clos_molon},
%but without expanding in ${u_0}$,
%which does not allow to see its character as a finite-size scaling law.
%%FALTA COMPROBAR QUE ES LA MISMA QUE REDNER!!!
%%OJO, NO, REDNER NO DESARROLLA EN $l_0$!!!

%MOVER EL PARRAFO DE ARRIBA A LA SECCION DE BRANCHINGS!!!

%Although our expressions look different than
%those in Ref. ???, they are totally equivalent.

\subsection{Reflecting boundary at $x=\ell$}

We now consider first-passage times to $x=0$, starting from a
reflecting boundary at $x=\ell$. The initial condition is most
conveniently handled by injecting into an empty interval a
single particle at $x=\ell$. Together with a zero flux condition at
%the reflecting boundary $x=\ell$, 
this boundary for $t>0$,
this stipulates that
\begin{equation*}
j_{tr\ell}(t)= 
\left[v c(x,t) -D \frac{\partial c(x,t)}{\partial x}\right]_{x= \ell}=-\delta(t),
\end{equation*}
where the $-\delta(t)$ term ensures that the injection (hence minus sign) occurs at
$t=0$. In Laplace space, this boundary condition reads
\begin{equation*}
v c(x= \ell,s) -D {c'(x= \ell,s)}=-1.
\end{equation*}
Note that in this case there is no dependence on $u_0$, since the
particle is injected precisely at $x=\ell$. The absorbing boundary
condition at $x=0$ remains unchanged.

The solution of the diffusion equation (Eq.~\eqref{diffeqlaplace}) with these boundary
conditions is provided in detail in
Refs.~\cite{Redner} and \cite{Font_clos_molon}.
%TAMBIEN EN REDNER, 2.2.30, PERO CON EL SIGNO DE V CAMBIADO!!!!
%CAMBIAR REF FOR TRANS!!!
The outflux at the absorbing boundary is
\begin{equation}
f_{tr}(s)=
j_{tr0}(s)=
D c'(x=0,s) = \frac{e^{-P_e} R}
{ R \cosh R -P_e \sinh R },
%D c'(x=0,s) = \frac{e^{-P_e} R}
%{ \sqrt{P_e^2 +\tau s} \cosh \sqrt{P_e^2 +\tau s} 
%-P_e \sinh \sqrt{P_e^2 +\tau s} },
\label{jtrftr}
\end{equation}
where the subscript $tr$ denotes \textit{transmission} from the
reflecting to the absorbing boundary \cite{Redner}. In contrast to the
case of two absorbing boundaries, this Laplace transform is not even
in $P_e$. The exact moments are given by
%Expanding for $z
%= \tau s /P_e^2 \ll 1$ the moments are obtained,
\begin{align}
\langle t_{tr} \rangle &= \frac {\tau}{4 P_e^2} (e^{2 P_e} - 2 P_e -1), 
\label{tref} \\
\langle t_{tr}^2 \rangle &= \frac {\tau^2}{8 P_e^4} [
e^{4 P_e} + (1-6 P_e) e^{2 P_e} 
%- 
+
2 (P_e^2 -1) ], 
\label{tref2}
\end{align}
% este ultimo no me coincide con Wolfram, el resultado es de Wolfram.
% repasar que sea signo + en las notas!!!! OK!!
see Ref. \cite{Redner}.
%where the subscript $ref$ refers to the fact that we are dealing with the time
%from the reflecting to the absorbing boundary.
For the critical case ($P_e=0$), we recover
$$%\begin{equation}
f_{tr}^*(s)=
j_{tr}^*(s)=
%D c'(x=0,s) = 
\frac 1 {\cosh \sqrt{\tau s}}
$$%\end{equation}
as in Ref.~\cite{Font_clos_molon}.

Since $t_\ell$ and $t_{tr}$ are independent random variables, the
total time $t_{\ell tr}=t_\ell+t_{tr}$ until absorption, having first
reached $x=\ell$ from the origin, Laplace transforms
as~\cite{Font_clos_molon}
\begin{equation}
f_{\ell tr}(s)=f_\ell(s) f_{tr}(s)=
\frac{R^2 e^{-P_e}\sinh P_e}
{ P_e R \sinh R\cosh R -P_e^2 \sinh^2 R }
+\mathcal{O}(u_0).
\label{laquefaltaba}
\end{equation}
%The first moment of this distribution is additive, 
%$ \langle t_{\ell tr} \rangle 
%=\langle t_\ell + t_{tr} \rangle = \langle t_\ell\rangle + \langle t_{tr} \rangle $, 
%but not the second moment.
%This is developed in the next subsections.
$ \langle t_{\ell tr} \rangle $ and 
the inverse Laplace transforms of $f_{\ell tr}(s)$ and $f_{\ell tr}^*(s)$
are shown in Figs. \ref {Fig_moments} and \ref{Fig_distributions}.

\subsection{Entire diffusion problem: moments}

We are now in a position to study the entire diffusion problem. The
quantity of primary interest is the first-passage time to $x=0$,
denoted $t_{r}$, where the subscript $r$ refers to the so-called
reflection mode~\cite{Redner}. This time is a mixture of two times:
$t_0$ (for realizations that do not reach $x=\ell$ before absorption
at $x=0$), and $t_{\ell
  tr}=t_\ell + t_{tr}$ (for realizations that do reach $x=\ell$ before
absorption at $x=0$). The weight of each time is given by $1$ (to
zeroth order in $u_0$) and
$C_\ell$ (from Eq. (\ref{CL})), respectively. Thus, to lowest
order in $u_0$, the expected value of $t_{r}$ is
\begin{align}
\langle t_{r} \rangle &= \langle t_{0} \rangle + C_\ell  { \left(
  \langle t_\ell \rangle + \langle t_{tr}\rangle \right)} \notag \\
%\label{sumofmoments}
&=
\left\{
\frac{1}{2 P_e \tanh P_e} \left(1 + P_e \tanh P_e  - \frac {P_e}{\tanh P_e}\right)
\right. \notag \\
&+
\left.
\frac{ e^{P_e} P_e }{\sinh P_e}
\left[
\frac 1 {2 P_e^2} \left(\frac{P_e}{\tanh P_e} -1\right)
+
 \frac {e^{2 P_e} - 2 P_e -1}{4 P_e^2} 
\right]\right\}
\tau {u_0}
 +\mathcal{O}({u_0}^2),
\label{sumofmoments}
\end{align}
which is plotted in Fig.~\ref{Fig_moments}.

For the second moment we find
\begin{equation}
\langle t_{r}^2 \rangle = \langle t_{0}^2 \rangle + C_\ell  { \left( \langle t_\ell^2 \rangle + \langle t_{tr}^2 \rangle  +2 \langle t_\ell \rangle \langle t_{tr} \rangle\right)},
\label{sumof2moments}
\end{equation}
to lowest order in $u_0$.  This result is due to the fact that the
moments of any order (with respect to the origin $t=0$) are additive in a
mixture of random variables (including the corresponding weights), but
not for the sum of independent random variables ($t_\ell$ and
$t_{tr}$), for which only the variances are additive.  The
particular form of $\langle t_{r}^2 \rangle$ can be obtained directly
from Eqs. (\ref{t02_noapp}), (\ref{tL2}), and (\ref{tref2}).  To first
order in $u_0$, $\sigma_{r}^2=\langle t_{r}^2 \rangle$.

\subsection{Possible order parameters}

Asymptotically, the first moment of $t_r$, Eq.~(\ref{sumofmoments}), behaves as
\begin{alignat*}{2}
\langle t_{r} \rangle &\sim \langle t_0 \rangle \sim 
\frac {\tau u_0}{2 |P_e|} =\frac {x_0}{|v|}&&\qquad \mbox{ for } P_e
\rightarrow -\infty, \\
\langle t_{r}^* \rangle &=
\left[ \frac 1 3 + 1\cdot \left(\frac 1 6 + \frac 1 2\right) \right] \tau u_0 
=\frac {\ell x_0} D
&&\qquad \mbox{ for } P_e =0, \\
\langle t_{r} \rangle &\sim C_\ell \langle t_{tr} \rangle \sim 
\frac {\tau u_0 e^{2 P_e}}{2 P_e} =\frac {x_0 e^{\ell v / D}}{v}
&&\qquad \mbox{ for } P_e \rightarrow \infty.
\end{alignat*}
as obtained immediately from Eqs. (\ref{t0_noapp}), (\ref{tL}),
(\ref{CL}), and (\ref{tref}). On the other hand, the order parameter considered in
Refs. \cite{Font_clos_molon,Botet}, $t_{\ell tr}= t_\ell + t_{tr} $,
behaves as
\begin{alignat*}{2}
\langle t_{\ell tr} \rangle &\sim 
\frac {\tau}{|P_e|} =\frac {2\ell}{|v|}&&\qquad \mbox{ for } P_e \rightarrow
-\infty, \\
\langle t_{\ell tr}^* \rangle &=
\left(\frac 1 6 + \frac 1 2\right)  \tau
=\frac {2 \ell^2} {3 D}&&\qquad
\mbox{ for } P_e =0, \\
\langle t_{\ell tr} \rangle &\sim \langle t_{tr} \rangle \sim 
\frac {\tau e^{2 P_e}}{4 P_e^2} =\frac {D e^{\ell v / D}}{v^2}
&&\qquad \mbox{ for } P_e \rightarrow \infty,
\end{alignat*}
%
%EXISTE EL LIMITE TERMODINAMICO????
%
Finally, the order parameter of
Refs. \cite{GarciaMillan,Corral_garciamillan} behaves as
\begin{alignat*}{2}
C_\ell &\sim 2 u_0 {|P_e|} e^{-2|P_e|} =\frac {x_0 |v|}{D} e^{-\ell |v|/D}&&\qquad
\mbox{ for } P_e \rightarrow -\infty, \\
C_\ell^* &= u_0
=\frac {x_0} {\ell}&&\qquad
\mbox{ for } P_e =0, \\
C_\ell &\sim 2 u_0 P_e =\frac {x_0 v}{D}&&\qquad
 \mbox{ for } P_e \rightarrow \infty.
\end{alignat*}

All three quantities, $\langle t_r \rangle$, $\langle t_{\ell tr}
\rangle$, and $C_\ell$, are reasonable candidates for order parameters
--- finding the order parameter of a phase transition is often not
obvious~\cite{Fisher_school}: citing J. P. Sethna, ``there is often
more than one sensible choice''~\cite{Sethna_book}.

Typically, one expects that in the thermodynamic limit ($\ell
\rightarrow \infty$) the order parameter goes to zero for $P_e < 0$
and scales with a power of $\ell$ for $P_e> 0$ (if the order parameter
is extensive).  This is not the case for $\langle t_r \rangle$ or
$\langle t_{\ell tr} \rangle$, and there is no simple rescaling to
bring about the desired behavior.  Instead, one could redefine the
order parameters somewhat artificially as $\ell^{-1} \ln (|v| \langle
t_r \rangle/x_0)$ and $\ell^{-1} \ln (v^2 \langle t_{\ell tr}
\rangle/D)$, which below and above the critical point behave as
intensive order parameters --- but they are not well defined at the
critical point.  $C_\ell$, meanwhile, undergoes a transcritical
bifurcation at the critical point \cite{Corral_Alseda_Sardanyes}, but
below the transition it goes to zero exponentially rather than as
a power law of $1/\ell$. A drawback of all three order parameters is
the lack of an associated variance that diverges at the critical
point.

\subsection{Finite-size scaling for the moments of the distributions}

The equations in the previous subsections show that, when
$u_0$ is small, all the first-passage times ($t_0, t_\ell, t_{tr}, t_{\ell
  tr}$, and $t_r$) obey finite-size scaling laws \cite{Privman}.
%%scaling is the same as $t_0$
%%and different from $t_\ell$ and $t_{tr}$ \cite{Font_clos_molon}.
%
%, i.e., as 
%Eqs. (\ref{scalingfort0}) - 
%%\ref{scalingfort02} 
%(\ref{la20}), (\ref{f0tscaling}),
%(\ref{scalingguay}) and (\ref{scalingguay2})
%(but with different scaling functions).
%%ESTO ULTIMO HABRA QUE MOVERLO!!!
%Going back to the case of a general P\'eclet number, 
%From the previous expressions 
%one can realize that the moments fulfill always finite-size scaling laws.
%
%
%
Indeed, starting with $t_\ell$,
\begin{align*}
\langle t_\ell \rangle &= {\tau} G_{ \ell 1} \left(P_e\right)=
 \frac{\ell ^2}{D} G_{ \ell 1} \left(\frac {\ell v}{2D}\right), \\
{\langle t_\ell ^2 \rangle} &= 
\tau^2 G_{ \ell 2} \left(P_e\right)=
\frac{\ell ^{4}}{D^2} G_{ \ell 2} \left(\frac {\ell v}{2D}\right),
\end{align*}
to zeroth order in $u_0$, where $G_{\ell 1}(P_e)$ and $G_{\ell 2}(P_e)$
are scaling functions completely determined by Eqs. (\ref{tL}) and (\ref{tL2}).
The same scaling holds for $t_{tr}$ and $t_{\ell tr}$ (but
with different scaling functions).
%The scaling functions $G_{01}(P_e)$, $G_{02}(P_e)$, $G_{ \ell 1}(P_e)$, etc., 
%are obtained directly from Eqs. (\ref{t0_noapp})-(\ref{t02_noapp}), (\ref{tL})-(\ref{tL2}), and
%(\ref{tref})-(\ref{tref2}).
%Using Eqs. (\ref{sumofmoments}) and (\ref{sumof2moments})
%allows one to conclude that 
%${\langle t_{r} \rangle}$ and ${\langle t_{r}^2 \rangle}$
%scale in the same way as ${\langle t_{0} \rangle}$ and ${\langle t_{0}^2 \rangle}$,
%respectively.
%
%
In contrast, 
%to the lowest order in ${u_0}$,
from Eqs. (\ref{sumofmoments}) and (\ref{sumof2moments}),
\begin{align}
\langle t_{r} \rangle &= {{u_0} \tau} G_{r 1} \left(P_e\right)=
 \frac{x_0  \ell }{D} G_{r 1} \left(\frac {\ell v}{2D}\right),
\label{scalingfort0} \\
{\langle t_{r}^2 \rangle} &= 
{u_0} \tau^2 G_{r 2} \left(P_e\right)=
\frac{x_0 \ell ^{3}}{D^2} G_{r 2} \left(\frac {\ell v}{2D}\right),
%=\sigma_0^2,
\label{scalingfort02}
\end{align}
to first order in $u_0$. The moments of $t_0$ share the same scaling
(but with different scaling functions).
%
%For comparison, $t_\ell$ and $t_{ref}$ scale in a different way,
%
%
%
%
For further comparison,
\begin{equation*}
C_\ell=u_0 G_c\left(P_e\right)=\frac{x_0} \ell G_c\left(\frac {\ell v}{2D}\right),
\end{equation*}
to lowest order in $u_0$. Note that the position of the critical point
does not shift with $\ell$: it remains at $P_e=0$ (or $v=0$) for finite $\ell$.

%ESCRIBIR COMO SON ASINTOTICAMENTE???

\subsection{Entire diffusion problem: distribution}

The Laplace transform of the probability density of the first-passage
time, $t_r$, can be written as the weighted sum
$$%\begin{equation}
f_{r}(s) = (1-C_\ell) f_0(s) + C_\ell f_\ell(s) f_{tr}(s) = j_0(s) + j_\ell(s) f_{tr}(s).
$$%\end{equation}
From Eqs. (\ref{laqnovalepaf}), (\ref{jL}), and (\ref{jtrftr}), we
thus obtain
\begin{equation*}
f_{r}(s)= j_0(s) + j_\ell(s) f_{tr}(s)= 
\frac
{e^{-P_e u_0}\sinh[(1-u_0)R] }
{\sinh R}
+
\frac{2 e^{-u_0 P_e} R \sinh(u_0 R)}
{ R \sinh(2 R) -2 P_e \sinh^2 R },
\end{equation*}
which, at the critical point $P_e=0$, reduces to
\begin{equation*}
f_{r}^*(s)= 
\frac
{\sinh[(1-u_0) \sqrt{\tau s}] }
{\sinh \sqrt{\tau s}}
+
\frac{2 \sinh(u_0 \sqrt{\tau s})}
{ \sinh(2 \sqrt{\tau s})}.
\end{equation*}
%In both cases, the first term gives the contribution of the particles
%that do not hit the $x=\ell$ boundary and the second term that of the
%other particles.
Expanding in $u_0$, we find (see Eqs. (\ref{firstordeninell}),
(\ref{jL}), and (\ref{jtrftr}))
\begin{align}
f_{r}(s)
&= 
1-\left(P_e+ \frac R {\tanh R} \right) {u_0} +
%\left(\frac{e^{P_e} P_e {u_0}}{\sinh P_e} \right)
\left( \frac{e^{P_e} R }{\sinh R} \right)
\left(\frac{e^{-P_e} R}
{ R \cosh R -P_e \sinh R } \right){u_0}
+\mathcal{O}({u_0}^2) \notag \\
&= 
1-\left(P_e+ \frac R {\tanh R} -
 \frac{2 R^2 }
{ R \sinh (2R) -2 P_e \sinh^2 R }  \right)  {u_0}
+\mathcal{O}({u_0}^2),
\label{Laplacetransformbuena}
\end{align}
valid for $u_0 R \ll 1$.  At the critical point ($P_e=0$),
\begin{align*}
f_{r}^*(s) &=
1- \frac { \sqrt{\tau s}} {\tanh  \sqrt{\tau s}} u_0  +
\left( \frac{  \sqrt{\tau s} }{\sinh  \sqrt{\tau s}} \right)
\left(\frac{1}
{ \cosh  \sqrt{\tau s} } \right) u_0 
+ \mathcal{O}(u_0^2) \\
&=
1- \left( 
 \frac { \sqrt{\tau s}} {\tanh  \sqrt{\tau s}} 
-\frac{  \sqrt{4 \tau s} }{\sinh  \sqrt{4 \tau s}} 
\right) u_0
+ \mathcal{O}(u_0^2).
%1- \frac { \sqrt{\tau s}} {\tanh  \sqrt{\tau s}} u_0  +
% \frac{  \sqrt{4 \tau s} }{\sinh  \sqrt{4 \tau s}} 
% u_0,
\end{align*}
Plots of the inverse Laplace transforms of $f_{r}(s)$ and $f_{r}^*(s)$
are shown in Fig. \ref{Fig_distributions}. Note that all the results
presented here for diffusion processes are valid for $x_0 \ll \ell$.
%EXPLICAR MEJOR EL MAPPING ABS-ABS Y EL CASO REF!!:

\subsection{Finite-size scaling for the distributions}

The Laplace transforms of the probability densities of $t_\ell$,
$t_{tr}$, and $t_{\ell tr}$ obey simple finite-size scaling laws
(to zeroth order in $u_0$),
\begin{equation}
f_\ell(s)= 
\hat F_\ell\left(\tau s, P_e \right)=
\hat F_\ell\left(\frac{\ell^2 s} D, \frac {\ell v} {2D} \right),
\label{la25}
\end{equation}
from Eq. (\ref{bis21}),
where the scaling function $\hat F_{\ell}$ is exactly known.
Again, 
$f_{tr}(s)$ and $f_{\ell tr}(s)$ scale in the same way as $f_\ell(s)$
(with different scaling functions).
Inverting the Laplace transforms, we see that the probability densities 
%of $t_\ell$ and $t_{tr}$ 
also obey simple finite-size scaling laws for fixed $P_e$,
%called finite-size scaling laws,
\begin{equation}
f_\ell(t)= \frac 1 \tau F_\ell\left(\frac t \tau, P_e \right)
=\frac D {\ell ^2} F_\ell\left(\frac {D t} {\ell ^2}, \frac {\ell v} {2D} \right),
\label{felltscaling}
\end{equation}
with $f_{tr}(t)$ and $f_{\ell tr}(t)$ scaling in the same way.

In contrast, the Laplace transforms associated with $t_r$ and $t_0$
obey
\begin{equation}
%f_{r}(s)= \hat F_r\left( \tau s, {{u_0}^2\tau s},\frac {\ell v} {2D} \right),
f_{r}(s)= \hat F_r\left( \tau s, {u_0}, P_e \right)=
\hat F_r\left( \frac{\ell^2 s} D, \frac{x_0}\ell,\frac {\ell v} {2D} \right),
\label{la20}
\end{equation}
%with scaling function $\hat F_{r}$, 
to first order in $u_0$, from Eq. (\ref{Laplacetransformbuena}).
This is not finite-size scaling for fixed $x_0$, due to
the dependence on $u_0$; $f_0$ obeys an analogous equation.  The
corresponding densities obey
\begin{equation}
f_{r}(t)=\frac 1 \tau F_r\left(\frac t \tau,  {{u_0}},P_e \right)
=\frac D {\ell ^2} F_r\left(\frac {D t} {\ell ^2}, \frac {x_0} \ell,\frac {\ell v} {2D} \right),
\label{f0tscaling}
\end{equation}
with $f_{0}(t)$ scaling in the same way, but with its own scaling
function.  However, as the moments of $t_r$ and $t_0$ have an extra
factor $u_0$ in comparison with $t_\ell$ and $t_{tr}$, this suggests
that the densities and their transforms can be written with an extra
factor $u_0$ as well, i.e.,
\begin{equation}
f_{r}(t)=\frac {u_0} \tau F_r\left(\frac t \tau,  P_e \right)
=\frac {D x_0} {\ell ^3} F_r\left(\frac {D t} {\ell ^2},\frac {\ell v} {2D} \right),
\label{f0tscaling_bis}
\end{equation} 
where in a slight abuse of notation we have recycled the symbol $F_r$,
which now refers to a different scaling function. $t_0$ obeys
an analogous equation (now truly a finite-size scaling law).  The
%finite-size 
scaling law for $f_r(t)$ is compatible with the form
\begin{equation}
f_r(t) = \frac 1 m \left( \frac m t \right)^\alpha F\left(\frac t \tau , P_e\right)
\label{scalingguay}
\end{equation}
(and similarly for $f_0(t)$) with the new scaling function $F$ going to
a constant for small arguments and decaying very fast for large
arguments, and with $\alpha=3/2$ and $m={u_0}^2 \tau$ the minimum value
of $t$ (that is, for $t < m$ the probability density $f_0(t)$ can be
considered as zero).
%Indeed, Eq. (\ref{scalingguay}) with $\tau \gg m$ 
%implies that the moments scale as in Eq. (\ref{t0n}),
%see Refs. \cite{Christensen_Moloney,Corral_csf}, and 
This is in agreement with 
the power-law behavior shown in Eq. (\ref{powerlaw32}).
An alternative way to write the scaling law (\ref{scalingguay})
is:
\begin{equation}
f_r(t) = \frac 1 m \left( \frac m \tau \right)^\alpha \tilde F\left(\frac t \tau , P_e\right)
%=   \frac {m^{1/2} D^{3/2}} {\ell ^3}  \tilde F\left(\frac {D t} {\ell ^2}\right),
%=   \frac {D x_0} {\ell ^3}  \tilde F\left(\frac {D t} {\ell ^2},\frac{\ell v}{2 D}\right),
\label{scalingguay2}
\end{equation}
with the scaling function $\tilde F$ absorbing the power-law part with
exponent $\alpha=3/2$.  For $P_e=0$ this leads to the same scaling law
as in Ref. \cite{Corral_csf}.
%Note that Eqs. (\ref{felltscaling}), (\ref{scalingguay}), and (\ref{scalingguay2})
%constitute finite-size scaling laws.
%The same scaling that holds for $f_0(t)$
%holds also for $f_{r}(t)$.
In fact, that reference gives a more direct derivation of the scaling
laws (\ref{scalingguay}) and (\ref{scalingguay2}), but only for
$f_0(t)$ with $P_e=0$.  
%%Those results can be generalized to $P_e\ne 0$
%%just replacing all the dependence on $x$ there by $x-vt =x- 2 P_e \ell t/\tau$.

%%OJO, al hacer la mezcla de $f_0$ con la otra, se sigue cumpliendo el scaling 
%\ref{scalingguay}??

\section{Branching process and size distributions}

\subsection{Diffusion and random walks}

First, we recall the connection between diffusion and random walks.
Consider a random walk described by a position $X$ at time $T$, which
are both discrete and dimensionless. At each time step $T$, the
position $X$ increases by one unit with probability $q$, or decreases
by one unit with probability $1-q$.  The continuum limit of the random
walk is a diffusion process, with $x=X \delta_x$ and $t=T \delta_t$,
where $\delta_x$ and $\delta_t$ are elementary space and time units,
which tend to zero \cite{Redner}. The limiting process is described by
the diffusion equation, Eq.~\eqref{diffeq}, with $v=(2q-1)
\delta_x/\delta_t$ and $D=2q(1-q) \delta_x^2/\delta_t$.
%% Esto sale de comparar la solucion de la ec de dif sin C. contorno
%% con la sol de un RW, que se obtiene trivialmente de la distr binomial.
%%In the continuum diffusion limit both $\delta_x$ and $\delta_t$ will tend to zero,
%Note that $\delta_x^2/\delta_t$ has to be finite (to get a finite, non-zero $D$);
%therefore, in order to get a finite $v$ one needs to take $2q-1$ of the order $\delta_x$,
%i.e., $q$ very close to the critical point $q=1/2$.
In this limit, the results obtained for moments and probability
densities of diffusion processes are also valid for random
walks~\cite{Redner}.  As all the relevant equations of the previous
section can be written in terms of dimensionless quantities, we can
make the substitutions
\begin{equation}
v \rightarrow 2\left(q-\frac 1 2\right),
\hspace{1cm}
D \rightarrow 2q(1-q), 
\hspace{1cm}
\mbox{and} 
\hspace{1cm}
\tau \rightarrow \frac{L^2} {2q(1-q)},
\label{substi}
\end{equation}
together with $u_0=X_0/L$, with $X_0=x_0/\delta_x$ and
$L=\ell/\delta_x$; in particular
\begin{equation}
P_e=\frac {L(q-1/2)}{2q(1-q)}.
\label{bis60}
\end{equation}
As an illustration, Eq. (\ref{sumofmoments}) becomes
\begin{equation}
\langle T_r \rangle = 
%\frac 1 {4 P_e} \left(\frac{e^{3P_e}-\cosh P_e}{\sinh{P_e}} -3 \right) \frac{X_0 L }{2 q (1-q)} =
\frac 1 {4 P_e} \left(\frac{e^{3P_e}}{\sinh{P_e}} -\frac 1{\tanh P_e} -3 \right) \frac{X_0 L }{2 q (1-q)} 
+\mathcal{O}(u_0^2).
\label{sumofmomentsbis}
\end{equation}
Care is required approximating the (dimensionless) probability mass
functions of random walk times $T$, with the probability densities
for diffusion times $t$, which have the dimensions of time$^{-1}$. Note
that, in the former case, first-passage times are discretized in steps of 2
(since boundaries can only be reached in either an odd or even number of steps). 
Thus, the two functions are related via %the change of variables
\begin{equation}
f_T(T)=2 \delta_t f_t(T\delta_t),
\label{p16}
\end{equation}
where $f_T$ denotes the probability mass function for the random walk
and $f_t$ the probability density for the diffusion process.
The extra factor 2 with respect a standard change of variables
comes from the discretization of $T$.
In this way, rewriting Eqs. (\ref{felltscaling}) and
(\ref{f0tscaling_bis}) for random walk scaling laws, we obtain  
$$
f_L(T) 
%=\frac 1 \tau F_\ell\left(\frac t \tau, P_e \right)
%=\frac D {\ell ^2} F_\ell\left(\frac {D t} {\ell ^2}, \frac {\ell v} {2D} \right),
=\frac {4q(1-q)} {L^2} F_\ell\left(\frac {2q(1-q) T} {L ^2}, P_e \right)
\simeq \frac 1 {L^2} F_\ell\left(\frac {T} {2 L ^2}, P_e \right),
$$
with an analogous expression for $f_{TR}(T)$,
while
$$
f_{R}(T)
%=\frac {u_0} \tau F_r\left(\frac t \tau,  P_e \right)
%=\frac {D x_0} {\ell ^3} F_r\left(\frac {D t} {\ell ^2},\frac {\ell v} {2D} \right),
=\frac {4q(1-q) X_0} {L^3} F_r\left(\frac {2q(1-q) T} {L ^2}, P_e \right)
\simeq \frac { X_0} {L^3} F_r\left(\frac {T} {2 L ^2}, P_e \right),
$$ where we have used $q\simeq 1/2$ close to the critical point. 
The scaling functions $F_\ell$ and $F_r$ are the same as for
the diffusion process of the previous section.

\subsection{Branching processes}

We now consider the Galton-Watson branching process associated with
the random walk. The branching process starts with one single
member (also known as the root), defining the first generation. It
produces a random number of offspring drawn from a geometric
distribution, i.e. the probability of $k$ offspring is $(1-q)q^k$, $k
= 0, 1, 2,\ldots$, where $1-q$ is the success probability. Each of these
second generation offspring produce their own (third generation)
offspring, and so on, independently and identically.
%Note that one does not need to consider a time variable,
%although the number of generations can play that role.
The process can be visualised as a rooted tree.
%(note that for us the tree is the set of connected occupied vertices
%and not the set of all possible vertices, occupied or not, which would
%have an infinite coordination number). %(which is also a tree)].
In principle, $q$ is the only parameter of the model, but one can
introduce finite-size effects by stopping the branching process at
generation $L$.% in other words, the elements of this generation are
%special and have no offspring (as if $q=0$ for them, in contrast to
%the previous generations).

%\footnote{
%DECIR EN ALGUN SITIO QUE LOS TREES SON CLUSTERS, O AL REVES!!\\
%HACER UNA TABLA DE CORRESPONDENCIAS???
%}

The size $S$ of the branching process, or the size of the tree or
cluster, is given by its total population (total number of
offspring plus root, or, equivalently, the total number of vertices in
the tree) \cite{Harris_original,Corral_FontClos}.  In a finite system,
clusters can be classified into two types: percolating clusters, which
reach generation $L$,
%(with the root at the
%first generation)
%, i.e., $X=1$),
and non-percolating clusters, which do not.
We denote the size of each of these by the random variables
$S_{perc}$ (for percolating) and $S_{int}$ (internal, for non-percolating).
Overall, $S$ is a mixture of $S_{perc}$ and $S_{int}$.
%Obvioulsy, $S$ is a mixture of both
%(and does not have to be confused with the Laplace variable, $s$).

%Note that percolating clusters in the geometric Galton-Watson process
%correspond to ``genuine'' percolation, in the sense that there is a
%process that fully propagates through a system, although, in contrast
%to the case when the offspring distribution is binomial, here there is
%not a fixed underlying tree, with a well-defined coordination number,
%as the possible number of branches from a vertex is not bounded.
%Alternatively, one may think in an underlying tree with a coordination number
%which is infinite, and where the occupation of the sites ??
%is not done independently with a fixed probability (leading to a binomial distribution)
%but a dependence is introduced in the sense that when the first time an empty link?? appears,
%all the rest (which are infinite) become empty.

\subsection{Mapping from branching processes to random walks}

\begin{figure}[ht]
\includegraphics[width=.99\columnwidth]{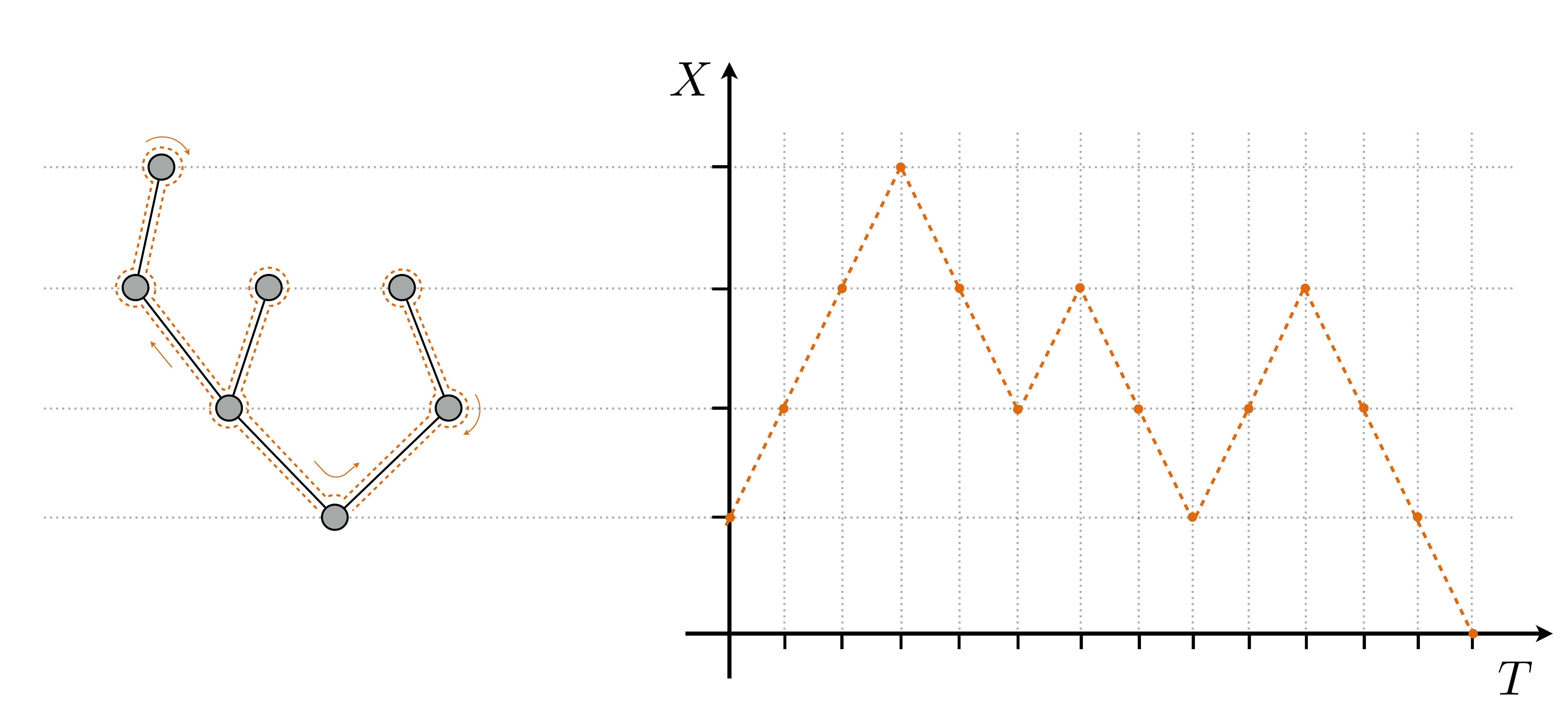}
\caption{ Correspondence between trees and positive walks
  (excursions), known as a Harris walk~\cite{Harris52}.  A walker
  starts at the root of the tree (the vertex at the bottom) and
  follows each branch in turn to its end, starting with the left-most
  branch. If the tree is generated from a geometric offspring
  distribution, the resulting path is a simple random walk.  }
\label{fig_mapping}
\end{figure}

Harris' mapping from trees to walks proceeds as follows (for complete
details see
Refs. \cite{Harris52,Font_clos_molon,Corral_garciamillan,Devroye_notes},
for a visual explanation, see Fig. \ref{fig_mapping}). A
(deterministic) walker is placed at the root of the tree and carries
out a so-called depth-first search by traversing each branch in turn
to its very end, starting with the leftmost branch. Whenever a choice
of unvisited branches presents itself, the walker traverses them in
the order left to right. Eventually, the walker will return to the
root, having traversed all branches. In doing so, the walker will have
visited each member of the tree twice. To define a positive
walk, it is convenient to append a final step to a ``generation
zero'', see Fig. \ref{fig_mapping}. In this way, one obtains a
one-dimensional, positive walk (or excursion), starting at $X=1$ and
ending at $X=0$, where $X$ corresponds to the generation number as the
walker traverses the tree. The size of the tree
$S$ is then seen to be the duration of the walk (plus one) divided by
two --- the division by two takes care of the fact that each member of
the tree is visited twice.

In addition, a probability measure over the set of all positive walks
is inherited from the probability measure over the set of rooted
trees.  When the tree is generated with a geometric offspring
distribution with success probability $p$, one obtains the standard
random walk, where $X\rightarrow X+1$ with probability $q=1-p$, and
$X\rightarrow X-1$ with probability $p$.

In this way, a realization of a Galton-Watson process with geometric
offspring distribution is equivalent to a realization of a
random walk that starts at $X=1$ and ends at $X=0$, staying positive
in between. The stipulation that a branching process cannot exceed $L$
generations is effected, in the random walk, by a reflecting boundary
at $X=L$.

%EL PARRAFO SIGUIENTE PARECE MUY REPETITIVO:
%The mapping between branching processes and random walks 
%\cite{Harris52,Font_clos_molon,Corral_garciamillan} establishes
%that a branching process with a limited number of generations
%corresponds to a random walk starting at $X=X_0=1$ and ending at $X=0$.
%Note that $X$ is discrete here. 
%If the branching process reaches the maximum number of generations,
%which is $L $, one has a percolating cluster, 
%corresponding to a random walk that finds a reflecting boundary at $X=L $
%and then ends at $X=0$.
%On the contrary, if the maximum number of generations is not achieved, 
%the cluster is non-percolating, and the random walker ends at $X=0$
%without hitting the position $X=L $. 
%
%PASAMOS DE rw A DIFUSION!!!

\subsection{Moments and splitting probability}

One can calculate $S_{perc}$ from the first-passage time of a
diffusing particle that first reaches $x=\ell$, before being absorbed
at $x=0$ (as in Ref. \cite{Font_clos_molon}). The total time is the
sum of two independent times: the first, $t_\ell$, is the time to
reach $x=\ell$ starting from $x=x_0$ (and not touch
$x=0$), and the second, $t_{tr}$, is the time to reach $x=0$ starting
from $x=\ell$, see Ref. \cite{Font_clos_molon}. Thus, in terms of a
random walk, $S_{perc}=(T_L+T_{tr}+1)/2$, where the discrete and
dimensionless times $T_L$ and $T_{tr}$ are analogs of $t_\ell$ and
$t_{tr}$ for the diffusion process. Similarly, $S_{int}$ is obtained
from the first-passage time to $x=0$ of a diffusing particle that does
not reach $x=\ell$, denoted $t_0$. Thus, $S_{int}=(T_0+1)/2$, with
$T_0=t_0/\delta_t$.  The total size $S$ (percolating or not) can be
obtained directly from the total first-passage time $t_{r}$, via $
S={(T_{r}+1)}/ 2, $ with $T_r=t_r/\delta_t$.
The moments of $S$ are
\begin{align*}
\langle S \rangle &= \frac{\langle T_{r} \rangle +1} 2 
%\simeq \frac{\langle T_{r} \rangle} 2
\simeq \frac{\langle t_{r} \rangle} {2\delta_t}, \\
%+ \frac 1 2,
\langle S^2 \rangle &= \frac{\langle T_{r}^2 \rangle +2  \langle T_{r} \rangle +1} 4
%\simeq \frac{\langle T_{r}^2 \rangle} 4
\simeq \frac{\langle t_{r}^2 \rangle} {4\delta_t^2},
%+\frac{\langle t_{r} \rangle} {2\delta_t} + \frac 1 4.
\end{align*}
%where we have used the large$-L$ limit and that $\langle T_{r}^n \rangle$
%scales as $L^{2(n+1-3/2)}$. 
for $L \gg 1$.
%$u_0\ll 1$.  
Thus, from Eq. (\ref{sumofmomentsbis}) 
%[which comes from Eq. (\ref{sumofmoments})],
$$
\langle S \rangle = 
\frac 1 {8 P_e} \left(\frac{e^{3P_e}}{\sinh{P_e}} -\frac 1{\tanh P_e} -3 \right) 
{ 2L } %{2 q (1-q)}
+\frac 1 2,
$$
%%{\bf This EQUATION is NEW!!!!}
%taking $X_0=1$ 
%(a direct imposition from Harris' mapping, as the branching process 
%starts with one single element)
where we have used the fact that, close to the critical point, $q\simeq 0.5$.
In the limit of large system size, the moments of $S$ obey the same scaling laws
as those of $t_{r}$, 
Eqs. (\ref{scalingfort0}) and (\ref{scalingfort02}), i.e.
%If we write $\ell$, $D$ and $v$ in their discrete versions, Eq. (\ref{substi}),  we
%get $$2 P_e\simeq 4 L \left(q-\frac 1 2\right) \simeq L  (\langle k \rangle -1)$$
%to first order in $q-1/2$, 
%with $\langle k \rangle = q/p \simeq 1+4(q-1/2)$
%for the geometric distribution. 
\begin{equation*}
\langle S \rangle = 
%{{u_0} \tau} G_{r 1} \left(P_e\right)=
\frac{  L }{4q(1-q)} G_{r 1} \left( \frac {L(q-1/2)}{2q(1-q)} \right)
\simeq {  L } G_{r 1} \left( {2 L(q-1/2)} \right),
\end{equation*}
Figure \ref{Fig_moments} shows good agreement between theory and
realizations of branching processes from computer simulations.

Note that the scaling law for $C_\ell$, Eq. (\ref{CL}), derived in the
diffusion framework, is the same as that obtained for branching
processes in Refs. \cite{GarciaMillan,Corral_garciamillan}.  Indeed,
transcribing $\ell$, $D$ and $v$ into their discrete versions,
Eq. (\ref{substi}), we get 
$$
2 P_e=\frac{\ell v} D\simeq 4 L \left(q-\frac 1 2\right)
\simeq L (\langle k \rangle -1)
$$
to first order in $q-1/2$, with
$\langle k \rangle = q/p \simeq 1+4(q-1/2)$ for the geometric
distribution.  Substituting the previous expression for $2P_e$ into
Eq. (\ref{CL}), with $x_0/\ell =1/L$, leads to the result of
Refs. \cite{GarciaMillan,Corral_garciamillan}.

\subsection{Distribution of sizes}

%In the same way we can calculate the distribution of $S$,
%starting with its Laplace transform,
%\begin{equation}
%f(S)=f_{int}(S) + C_\ell f_{perc}(S) =
%\end{equation}
%
%$$
%f_{int}(S)=
%% f_T(2S-1) =
%\frac 2 {\delta_t} f_0((2S-1)\delta_t) 
%$$

The distributions of sizes $f_S(S)$ (for $S_{int}, S_{perc}$
or total $S$) are related to the distributions of first-passage times in
the random walk $f_T(T)$ and in the diffusion process $f_t(t)$ via
\begin{equation*}
f_S(S)=
f_T(2S-1)=2 \delta_t f_t((2S-1)\delta_t),
\end{equation*}
using Eq. (\ref{p16}). While we do not have explicit formulas for
$f_T(T)$ and $f_t(t)$, we do have their Laplace transforms $f_t(s)$.
Thus, given a finite-size Galton-Watson branching process, with
parameters $q$ and $L$, we can calculate $\tau$, $P_e$, and $u_0$ for
the equivalent diffusion process, using Eqs. (\ref{substi}) and
(\ref{bis60}), and then perform the numerical inversion of the
Laplace-transformed expressions (\ref{nuevanueva_noapp}),
(\ref{laquefaltaba}), and (\ref{Laplacetransformbuena}).  This
yields a nearly perfect agreement between computer
simulations of the Galton-Watson process and theory, based on
diffusion processes, as Fig. \ref{Fig_distributions} illustrates.

Note from the figure that the critical case, $P_e=0$, displays a bump
before the exponential decay at large sizes, which comes from the
Kolmogorov-Smirnov distribution associated with percolating clusters.
Similar bumps have been observed in paradigmatic models of critical
phenomena, such as the Oslo sandpile model~\cite{Corral_thesis}.
Therefore, although deviations from criticality ($P_e\ne 0$)
in infinite systems lead to a $f(S)$ given by a power-law
multiplied by an exponential tail 
($f(S) \propto e^{-S/\xi}/S^{3/2}$, see Ref. \cite{Corral_FontClos}),
this parameterization is not valid for finite-size effects at the critical point,
%and above it, 
since it does not reproduce the bump for large $S$.
A much larger bump is present in the supercritical regime.

This behavior can be taken as an instance of the so-called dragon-king
effect \cite{Sornette_dragon_king}, in which events at the tail of a
distribution have a much larger probability than that given by the
extrapolation of a power-law central part.  This would correspond to
the so-called characteristic-earthquake scenario in statistical
seismology \cite{Ben_Zion_review}, although from our results and those
of Ref. \cite{Font_clos_molon} it seems clear that the bump cannot be
described by Gaussian-like statistics (which is only applicable in the
subcritical regime, where the bump is negligible or very small,
depending on $P_e$), contrary to the statement in
Ref. \cite{Ben_Zion_review}.

Finally, from Eqs. (\ref{felltscaling}) and (\ref{f0tscaling_bis}),
the scaling laws for the size distributions can be written as
\begin{align*}
f_{perc}(S)
&=\frac {1} {L^2} F_{\ell tr}\left(\frac {2S-1} {2 L ^2}, P_e \right), \\
%OJO!!! La func de escala no esta definida antes!!!!!
f_{int}(S)
&=\frac {1} {L^3} F_0\left(\frac {2S-1} {2 L ^2}, P_e \right),\\
f(S)
%=\frac {u_0} \tau F_r\left(\frac t \tau,  P_e \right)
%=\frac {D x_0} {\ell ^3} F_r\left(\frac {D t} {\ell ^2},\frac {\ell v} {2D} \right),
&=\frac {1} {L^3} F_r\left(\frac {2S-1} {2 L ^2}, P_e \right),
\end{align*}
where, again, $q\simeq 1/2$ close the the critical point.
%(a direct imposition from Harris' mapping, because the branching process 
%starts with one single element)
The scaling functions $F_\ell$, $F_0$, and $F_r$ are the same
as for the diffusion process.

\section{Summary and Conclusions}

The Harris walk mapping establishes a direct correspondence between
finite-size branching processes (in which the number of generations
cannot exceed $L$) and one-dimensional random walks between
$X=0$ and $X=L$.  By approximating random walks with diffusions,
techniques from the latter can be applied 
to branching processes, such that first-passage times of
Brownian particles with drift correspond to sizes of trees
generated by the branching process (up to a proportionality factor).

We solved the ensuing diffusion equations, arriving, in the limit of
large system size, at exact expressions for the Laplace transforms of
the probability densities of first passage times. The drift term is a
control parameter, bringing about a second-order phase transition as it
changes sign. This transition separates a regime in which diffusing
particles (starting at the origin) barely reach the distant boundary,
from a regime in which particles reach the boundary with non-zero
probability. In the context of branching processes, the transition
separates subcritical and supercritical phases. In the latter case,
trees percolate (i.e. reach generation $L$) with non-zero probability.
In the limit of infinite system size, these transitions are sharp.
We obtained finite-size scaling laws for probability densities, and
discussed possible choices of order parameter.

Our approach allows us to treat separately the contribution from
particles that do reach the further boundary (corresponding to
percolating trees) and particles that do not.  In the latter case, the
distribution is governed by a power law with exponent $3/2$ (except
for very large and very short times), whereas in the former case we
recover the results of Ref. \cite{Font_clos_molon}, which give a
Kolmogorov-Smirnov distribution in the critical case.

An important lesson from this study is that truncated gamma
distributions \cite{Serra_Corral} (power laws multiplied by an
exponential decay term), although valid for modeling off-critical
effects in infinite systems, are not appropriate for modeling
finite-size effects in critical systems, due to a large-size bump in
the distribution coming from system-spanning clusters.  Another point
to bear in mind is that the existence of finite-size scaling in the
distributions of some observable is not a guarantee that the system
under consideration is at a critical point. It could be that the
system is simply close to but not at the critical point, 
in such a way that the rescaled control parameter ($P_e$ in our case)
takes a constant value.
%If $v$ is
%close to zero but fixed, finite-size scaling is not observed.

%OJO AL CAMBIO DE VARIABLE, 
%HAN DE SALIR FACTORES 2 !!!!??

%Discusion: parametros de orden, citar Fisher, Sethna!!!!???

%\footnote{siempre salen las particulas que tocan la otra frontera???
%Si porque la distrib sale normalizada}

%\footnote{$s_{int}$ es simetrico respecto a la velocidad??}

%logarithmic binning \cite{Hergarten_book,Christensen_Moloney,Pruessner_book,Corral_Deluca}

%%citar Devroye!!!???

\section*{Acknowledgements}

We acknowledge
support from projects
%FIS2012-31324, 
FIS2015-71851-P, MAT2015-69777-REDT,
and the Mar\'{\i}a de Maeztu Program for Units of Excellence in R\&D (MDM-2014-0445)
from Spanish MINECO, 
as well as 2014SGR-1307, from AGAUR.
A.C. appreciates the warm hospitality of the London Mathematical Laboratory.

\section{Appendix A}
We provide details of the calculation of the outflux at $x=0$ in
a system with two absorbing boundaries, see Eqs.~(\ref{diffeq}),
(\ref{abc}), and (\ref{deltainitial}).  From Eq.~(\ref{lagorda}), we
arrive at the exact expression
\begin{equation}
j_0(s)=D c'(x=0,s)=\frac{e^{-P_e {u_0}}\sinh[(1-{u_0})R]}{\sinh R},
\label{laqnovalepaf}
\end{equation}
from which
\begin{equation*}
  f_0(s) = \frac{j_0(s)}{j_0(s=0)} =
  \frac {\sinh[(1-u_0)R] \sinh P_e}
        {\sinh[(1-u_0)P_e] \sinh R}.
\end{equation*}

We are interested in particles starting very close to the $x=0$
boundary, i.e.  $x_0 \ll \ell$ and ${u_0} \ll 1$. Expanding
Eq. (\ref{laqnovalepaf}) to first order in ${u_0}$,
\begin{equation}
j_0(s)=1-\left(P_e+ \frac R {\tanh R} \right) {u_0} + \mathcal{O}({u_0}^2)
\label{firstordeninell}
\end{equation}
%
%TENDRIAMOS QUE DISTINGUIR 2 CASOS AQUI, R PEQ Y R GRANDE, 
%O SEA TANH LINEAL Y CONSTANTE!!!! VER REDNER 49.
%O NO, mejor R TENDIENDO A Pe O A $\sqrt{\tau s}$.
%OTRO CASO ${u_0} R=\sqrt{\frac {x_0^2 v^2}{2^2D^2}+ x_0^2 s/D}$
%(no depende de L y es siempre pequenyo!!)!!
%
(in fact, we require $u_0 P_e \ll 1$ and $u_0 R \ll 1$).  The
properties of the first-passage time to $x=0$ will arise from the Taylor
expansion of $j_0(s)$ around $s=0$.  If we write $R=P_e
\sqrt{1+z}$ with $z=\tau s/P_e^2$, then, to second order in $z$,
%then, we first expand the square root 
%%in $R$ 
%and then apply the formula
%for the $\tanh$ of a sum, to finally obtain, up to second order in $z$
\begin{align}
  &j_0(s)=1-\left(P_e + \frac{P_e}{\tanh P_e}\right) {u_0} -
  \frac{P_e}{2 \tanh P_e}
  \left(1 + P_e \tanh P_e - \frac {P_e}{\tanh P_e}\right){u_0} z
  \notag \\
  &-\frac{P_e}{2 \tanh P_e}\left( 
  -\frac 1 2 + \frac {P_e \tanh P_e} 2  - \frac{P_e}{2 \tanh P_e}  -P_e^2 + \frac{P_e^2}{\tanh^2 P_e} 
\right) \frac{{u_0} z^2} 2 + \mathcal{O}(z^3){u_0} + \mathcal{O}({u_0}^2).
\label{j0lin}
\end{align}
Since $j(s)$ is the moment generating function for first passage
times, i.e. $j(s)= C (1-\langle t \rangle s + \langle t^2 \rangle
s^2/2 + \mathcal{O}(s^3))$, for some normalization constant $C$, we
read off from the above expansion
\begin{align}
\langle t_0 \rangle &=
\frac{1}{2 P_e \tanh P_e} \left(1 + P_e \tanh P_e  - \frac {P_e}{\tanh P_e}\right)\tau {u_0}
+ \mathcal{O}({u_0}^2),
\label{t0} \\
%CUANDO SE SUMA A tL, VERIFICAR QUE DA LO MISMO QUE REDNER 2.3.10, P 58!!??
% si, ver el apendice siguiente
\langle t_0^2 \rangle &=
\frac{1}{2 P_e^3\tanh P_e}\left( 
\frac 1 2
 - \frac {P_e \tanh P_e} 2  + \frac{P_e}{2 \tanh P_e} +P_e^2- \frac{P_e^2}{\tanh^2 P_e} 
\right) {\tau^2{u_0}}  + \mathcal{O}({u_0}^2),
\label{t02}
\end{align}
where the subscript $0$ in $t$ denotes first-passage to the boundary
at $x=0$.  To first order in $u_0$, the variance coincides with the
second moment, i.e., $\sigma_0^2=\langle t_0^2 \rangle - \langle t_0
\rangle^2 \simeq \langle t_0^2 \rangle$.

The zeroth-order term in $s$ (i.e., the constant $C=C_0=j_0(s=0)$), is
only one in the limit ${u_0} \rightarrow 0$.  To first order in
$u_0$, the coefficients in the expansion of $j_0(s)$ yield the moments
of $t_0$.  Using Eq. (\ref{firstordeninell}), the
Laplace-transformed probability density is
\begin{equation}
f_0(s)=
\frac{j_0(s)}{j_0(s=0)}=
1-
\left(\frac R {\tanh R} - \frac {P_e} {\tanh P_e} \right) {u_0} 
+ \mathcal{O}({u_0}^2),
\label{nuevanueva}
\end{equation}
to first order in $u_0$.

%\section*{Appendix II}
%Finally, let us note that the next formula, valid for $b \ll 1$ and $b
%\ll \tanh a$, has been useful in the previous calculations:
%$$
%\frac {a+b}{\tanh(a+b)} =
%$$
%$$%\begin{equation}
%\frac a {\tanh a} \left[ 1 + \frac b a + 
%\left(\tanh a -\frac 1 {\tanh a}\right) b +
%\left(\tanh a -\frac 1 {\tanh a}\right) \frac {b^2} a +
%\left(-1 + \frac 1 {\tanh^2 a}\right) b^2  
%\right].
%%+ \mathcal{O}(b^3).
%$$%\end{equation}
%\begin{equation}
%=
%\frac a {\tanh a} \left[ \left(1 + \frac b a \right) \left(1+b 
%\left(\tanh a -\frac 1 {\tanh a}\right) \right)+
%\left(-1 + \frac 1 {\tanh^2 a}\right) b^2  
%\right]
%%+ \mathcal{O}(b^3).
%\end{equation}

\section{Appendix B}

The problem of one-dimensional diffusion between two absorbing
boundaries (analyzed in Ref. \cite{Farkas}) displays a phase
transition in the same way as diffusion between absorbing and
reflecting boundaries. The calculation of first-passage times is
analogous to that of the absorbing-reflecting system, but with the
contribution from $t_{tr}$ excluded.

The exact Laplace-transformed probability density for $t_{0\ell}$, the
first-passage time to either boundary, reads
\begin{equation*}
  f_{0\ell}(s)= \frac{e^{-P_e {u_0}}\sinh[(1-{u_0})R]}{\sinh R} +
  \frac{e^{(1-{u_0})P_e} \sinh({u_0} R)}{\sinh R},
\end{equation*}
which, at the critical point ($P_e=0$), reduces to
\begin{equation*}
  f_{0\ell}^*(s)=\frac
  {\sinh[(1-u_0)\sqrt{\tau s}] + \sinh(u_0\sqrt{\tau s})}
  {\sinh\sqrt{\tau s}}.
\end{equation*}
These expressions may be expanded in $u_0$ as
\begin{equation*}
  f_{0\ell}(s)=j_0(s) + j_{\ell}(s) = 1-u_0
  \left(P_e+\frac R {\tanh R}-\frac{e^{P_e} R}{\sinh R}\right)
  + \mathcal{O}(u_0^2),
\end{equation*}
and, at the critical point,
\begin{equation*}
  f_{0\ell}^*(s) = 1-u_0
  \left(\frac{\sqrt{\tau s}}{\tanh{\sqrt{\tau s}}} -
  \frac{\sqrt{\tau s}}{\sinh{\sqrt{\tau s}}}
  \right) +\mathcal{O}(u_0^2),   
\end{equation*}
in which the Kolmogorov-Smirnov distribution again appears
(corresponding to particles that reach $x=\ell$). Note that
$t_{0\ell}$ scales in the same way as $t_r$ and $t_0$ (but with
different scaling functions).  Thus, the scaling laws in the
core of the paper also apply here (but with different scaling
functions).

To first order in $u_0$, $\langle
t_{0\ell}\rangle= \langle t_{0} \rangle + C_\ell \langle t_\ell
\rangle$, where $C_\ell$ is the same as in the absorbing-reflecting system.
Thus, the first moment is given by
\begin{equation*}
\langle t_{0\ell}\rangle=
\frac {\tau u_0} 2
\left(1+\frac 1 {\tanh P_e}- \frac 1 { P_e} \right),
\end{equation*}
which is the expansion to first order in $u_0$ of Eq. 2.3.10 in
Ref. \cite{Redner}, or, equivalently, Eq. (6) in Ref. \cite{Farkas}.

%\bibliographystyle{unsrt}

%%\bibliography{../../../words_ramon/p1_lemmas/biblio}
%%\bibliography{C:/Users/Alvaro/Dropbox/p1_lemmas/biblio} % portatil
%%\bibliography{C:/Users/acorral/Dropbox/p1_lemmas/biblio}  % fijo
%%\bibliography{C:/Users/acorral/Dropbox/p1_lemmas/biblio}
%%\bibliography{biblio} 

%\bibliographystyle{unsrt}
%\bibliography{biblio}

%\newpage
%{\tiny
%\begin{verbatim}
%1d diffusion <-> 1d RW <-> geometric BP
%                                 | 
%                            binomial BP  <-> mean-field   <-> 2nd order         <-> bifurcations in
%                                             percolation      phase transitions       dynamical systems
%                                         <-> SOC
%\end{verbatim}
%}

\newpage

\newpage

\newpage

\newpage

\end{document}